# Assistive Visual Cues for Visual Neglect Patients

**Per Bjerre**
Aalborg University
Denmark
pbjerr10@student.aau.dk

**Andreas Køllund Pedersen**
Aalborg University
Denmark
akpe10@student.aau.dk

**Hendrik Knoche**
Aalborg University
Denmark
hk@create.aau.dk

**ABSTRACT**

Previous research on exogenous and endogenous cues has shown how they direct attention and improve interaction speed and error rate in applications. However, most studies focus on people with normal sight. People suffering from visual neglect have difficulties attending to parts of the visual field. One treatment method calls for the use of strong visual cues to remind patients of their neglected area and help guide their attention to it. Therefore, we examine the effects of endogenous and exogenous cues on visual neglect patients. Our results showed that visual neglect patients perform better with endogenous cues, when targets are within their neglected area. In some cases, combining exogenous and endogenous cues improve performance further. However, the performance varies greatly between patients. Using one neglect patient as an example, we saw that the best endogenous cue had an average acquisition time of 3.5 seconds compared to 6.5 for the best exogenous. Combining exogenous and endogenous cues further improved acquisition time to 2.8 seconds.

**Author Keywords**

Visual Search; Visual Cues; Visual Feedback; Visual Neglect;

## 1. INTRODUCTION

Cues designed to guide and direct attention are everywhere around us, such as the brake lights of a car driving in front of us or the light on the toaster calling attention to the fact that it is turned on. Guiding attention through cues is a powerful assistive tool and is an area that rapidly expands as technology and availability allows it. However, some individuals have difficulties registering these visual inputs. An extreme example of this is visual neglect. Neglect patients have difficulties performing everyday tasks, such as reading and grooming. In severe cases, they can even have trouble walking through a door.

Previous research on assistive and treatment tools for visual neglect has shown that sensory cues can improve patient's ability to do everyday tasks and even reduce the size of the neglected area [3, 11]. However, we found little research on how visual cues influence the search patterns of visual neglect patients and whether they can respond to stimuli positioned within the neglected area or extending into the area.

We investigate how visual cues influence visual search patterns, acquisition rate, and acquisition time for patients suffering from visual neglect. Furthermore, we compare the findings to young and middle-aged participants. First, we examine previous research on visual neglect along with diagnostic and treatment methods, how cues can guide attention and examples of assistive cues. The goal is to investigate how visual cues can assist patients suffering from visual neglect in virtual search tasks.

## 2. BACKGROUND

### 2.1 Visual Neglect

Visual neglect, also known as unilateral- or hemispatial neglect, is an attention deficit in which patients ignore or do not respond to stimuli on one side of the visual field. While the deficit can occur in any area of the visual field, it often manifests itself as either a right or left vision deficit. The primary cause of visual neglect is damage to the cerebral hemispheres of the brain, such as a stroke. [3, 5]

Pencil and paper tasks requiring visual accuracy can help assess visual neglect. These tasks include trail making, line bisection, and cancellation tasks. See Figure 2.1 for an example of a trail making task. These tasks can quantify the extent and type of neglect, as the boundary between neglected and non-neglected space may not be a straight line on the horizontal plane, but rather a gradient that varies in spatial location from left to right. [3, 5]

Cherney [3] defined three basic approaches to treating visual neglect. The first approach adapts the environment to help patients cope with the neglect, by reducing the patient's dependence on the impaired process. The treatment includes placing non-subtle cues in the environment, such as having brightly colored items on the left side of the cupboard, to help the users through daily activities. The second approach involves repeatedly performing specific everyday tasks to gain more independence in doing them.

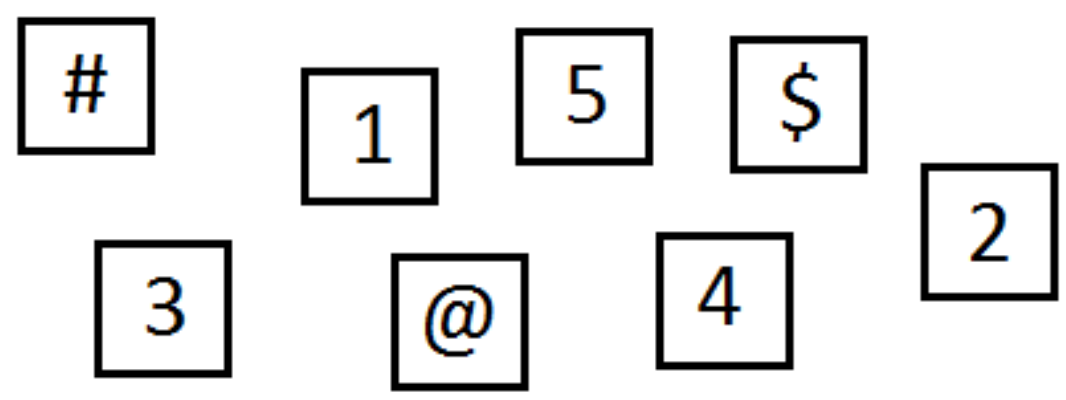

**Figure 2.1: A trail making test. The user connects the numbered targets in an ascending order. The symbols are additional distracters.**

The third approach uses the hypothesis that the patients suffering from neglect do not move their eyes to attend to the neglected side automatically, unless they get explicit cues to do so. Techniques based on this approach train the user to pay attention to the neglected side through explicit cues, drawing the user's attention to that side. For example, briefly touching the patients left arm, to draw attention to the left side. [3]

#### 2.1.1 Visual Restoration Therapy

In the area of visual neglect, Visual Restoration Therapy (VRT) has been used a treatment method to improve the useable visual field of neglect patients. VRT uses the third approach described by Cherney [3] by training patients to attend their neglected area. The basic principle of VRT is to assess the size and location of the neglected area, then present the patient with visual cues along the borders of the neglected area. The cues repeatedly remind patients to attend to the neglected area, and this effect should shrink the area over time. VRT uses reaction time and acquisition rate (defined as the amount patients detected a visual stimulus), within the neglect area, to determine to effectiveness of the treatment. [11] This methodology is applicable to determine the effectiveness of visual cues.

### 2.2 Guiding Attention

This section examines visual feature channels; along with how they help users differentiate between targets. Feature channels describe physical features of objects, such as size, shape, and color.

#### 2.2.1 Attention Guiding Attributes

To understand the basic components of visual attention, we explore the visual cuing paradigm [14]. The visual cueing paradigm describes two types of cues, exogenous and endogenous cues, used to find a visual target among distracters. Exogenous cues subconsciously attract attention, for example by altering the size or color [15]. These cues start to attract visual attention instantaneously, but the attention fades after 200-600ms. Endogenous cues require the users to consciously direct attention towards a target, for example by pointing towards the target. These cues attracts attention slower, become stronger than exogenous cues after 300ms. [14]

Wolfe and Horowitz [16] compiled a meta-analysis of different visual feature channels and organized them into categories, based on the strength of the feature channel. They based the distribution on reaction time and converging evidence from existing research on the specific cues. The features of color, motion, orientation, and size have the highest probability of catching attention. Less probable feature channels include luminance pulsations, shape, and expansion. [16]

Ware [15] described how exogenous cues can reduce visual search time by subconsciously leading gaze towards selected regions. The use of distinctive feature channels can help targets stand out from their surroundings. An example of this is a colorful target on an otherwise black and white background. Targets should differ in more than one feature channel to further decrease search time. [15]

For touch interfaces, Benko and Wigdor [1] used exogenous motion cues to alleviate target occlusion. For successful selections, an inwards contracting circle would appear around the touch point. For unsuccessful selections, an outwards expanding circle around the touch point appeared instead. These circles had a diameter of 55 mm. They used these cues to give visual feedback on whether users hit a target or not. [1]

The positioning of cues in the visual field also influences performance. In comparison to the center of gaze, the peripheral vision is highly sensitive to movement, however poor at distinguishing color and shape. This is due to a lower density of receptors for the latter cue types in peripheral vision. [7] Further, McColgin [10] found that there is no difference between clockwise and counterclockwise movement in peripheral vision, and that it is easier to perceive vertical movement compared to horizontal [10]. The Useful Field of View (UFOV), described by Ware [15], is a region of the visual field that processes information quickly. The size of the UFOV have been shown to vary greatly dependent on the density of symbols or objects within an area, going from a 1–4-degree angle for high density up to 15 degrees for low density areas. In this case, they classified low density as less than one target of 2.54 mm$^2$ per degree of visual angle. The cognitive load of a task also influences the performance of search, with a drop from 75 to 36 % in detection rate of peripheral cues when performing tasks requiring high amounts of attention. This means that if the task requires high cognitive load, the peripheral cues should be stronger than usual. For movement in particular, people can respond within one second to moving targets within 20 degrees of the fixation point, before performance declines. For static targets, response time increases rapidly from below one second to 5-110 seconds, after crossing four degrees from the fixation point. [15]

Gustafson et al. [6] proposed an endogenous wedge-based visualization technique that conveys both direction and distance to off-screen targets. Users can visually trace the two legs of a wedge beyond the borders of the screen and estimate where they intersect. Participants were significantly more accurate (27 %) when using the wedge technique than when using a similar halo-based technique. They also noted that both techniques were equally good at conveying distance. Based on this principle, the technique should also be able to convey information about targets located within the neglected space, if the wedge extends into the non-neglected area, of visual neglect patients. [6]

In the area of augmented reality, Biocca et al. [2] investigated different approaches to guide attention for search tasks in a 360-degree omnidirectional workspace.

They examined exogenous highlighting of targets, auditory position cues through headphones, and another type of visual feedback called an attention funnel. The attention funnel worked by superimposing an endogenous tunnel for the user to look through in order to find the target. The tunnel curved towards the target to illustrate direction. This way, the users get a visual direction to the target regardless of the orientation of the user in relation to the target. The study showed that the funnel approach had a shorter search time of 44.7 seconds compared to the 65.5 seconds of exogenous highlighting. They reported no significant difference in error rate between the two approaches. [2] An advantage of using a funnel is that users are guided towards a target regardless of their orientation to the target. The funnel could benefit from knowing the fixation point of users, to redirect the gaze to a specific target.

#### 2.2.2 *Attention Cues in Eye Tracking*

Majaranta et al. [9] investigated the effect of different visual feedback cues on gaze controlled typing. For visual feedback, they used a exogenous red color change to indicate selections. The feedback improved interaction speed and acquisition rate by 10.3 %. They found that using exogenous visual cues off target increased input and error rate, as off target exogenous cues direct attention away from the interaction target. However, when purposely guiding attention elsewhere, this could be a benefit instead. [9, 12]

## 3. METHODS AND MATERIALS

In this section, we explore the use of visual cues to assist users in directing their attention towards specified targets. The first study focuses on the reaction time of exogenous cues. In the second study, we examine the use of endogenous cues to direct attention into the neglected area through visual search tasks. The third study extends upon the findings from the previous studies and combines exogenous and endogenous cues.

### 3.1 First Study

Exogenous cues are the basis for one of the treatment approaches for visual neglect [3]. Therefore, we conducted a study that focused on the use of exogenous within the neglected area of patients. The study used reaction time and acquisition rate to determine the performance of the cues. We used four strong visual cues, classified by Wolfe and Horowitz [16]: object expansion, luminance pulsation, peripheral movement, and color (denoted as *expansion*, *luminance*, *peripheral motion*, and *color*). See Figure 3.1 for an illustration of the cues.

#### 3.1.1 *Design and Procedure*

The study used a within subject design, with the endogenous cues as the main independent variable. Further, we examined the effect of target distance from the center. The dependent variables were reaction time from each stimulus onset and acquisition rate, with acquisition rate

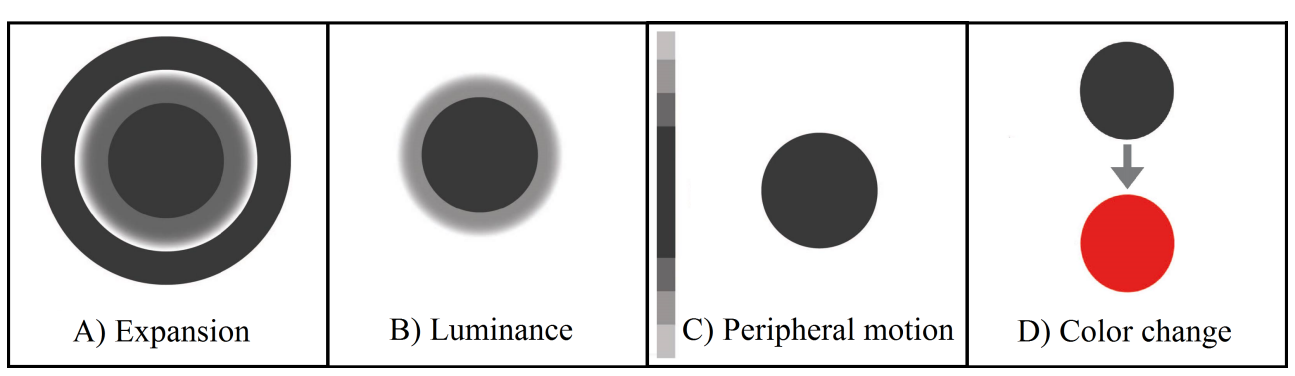

**Figure 3.1: Illustration of the four exogenous cues for the first study. (A), (B), and (C) iterated in stimuli.**

being the amount of time participants responded to the stimuli.

The task consisted of responding to visual cues appearing on one of the targets in the scene, by pressing the 'space' key as soon as they noticed the cue. At full luminance, *luminance* had a contrast ratio of 3:1 from the initial color. A cycle of the cue lasted 1.2 seconds. *Color* changed from gray to red and had a contrast ratio of 3.2:1. The *peripheral motion* bar had a height of 20 cm and width of 0.5 cm. An iteration of the cue lasted 50 ms (moving from one side to the opposite). An iteration of *expansion* lasted 30 ms and pulsated between 4 and 16 mm (400% size increase). The cues appeared in a randomized order and each cue appeared for three seconds or until 'space' was pressed. The next cue would appear between one and two seconds after the previous cue had disappeared. Cues appeared on targets of 6 mm in diameter, placed in a grid structure covering the whole screen, similar to the grid approach used for VRT [11]. The grid had five columns and three rows on each side of the centerline, with 30 targets total. The columns had a distance of 4, 8, 12, 16 and 20 cm to the centerline. With participants at an average visual distance of 50 cm, this results in visual angles of 4.6, 9.1, 13.5, 17.7 and 21.8 degrees from the centerline. The center of the screen had a cube, which users had to fixate on during the test. This cube had a diameter of 20 mm in diameter and gradually changed color to keep attention. The duration of one cycle was 2 seconds. See Figure 3.2 for an overview of the grid setup and Figure 3.3 for the setup.

#### 3.1.2 *Participants*

For the experiment, we had access to two patients with visual neglect, with the ages of 21 and 72. The study included three patients with an attention deficit, between the ages of 23 and 71 ($M$ = 54.33, $SD$ = 27.15), and three voluntary students from the local university campus, in the age range of 24 to 27 ($M$ = 25.33, $SD$ = 1.53).

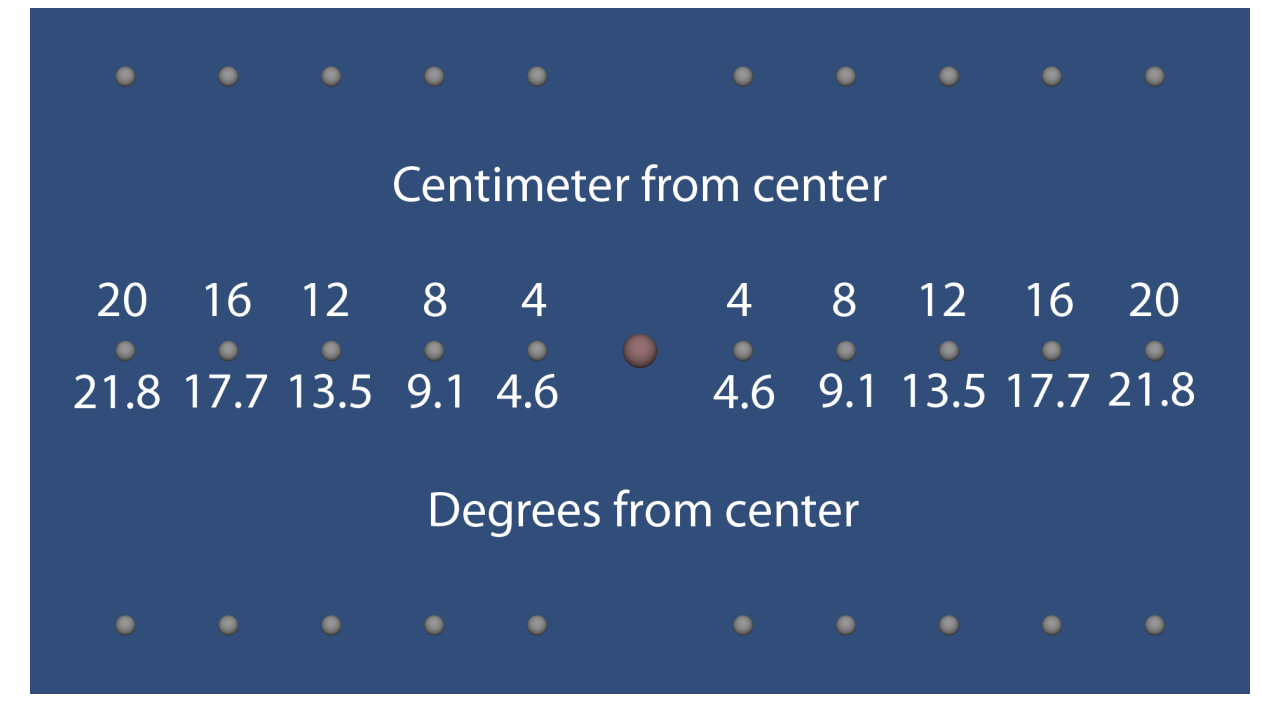

**Figure 3.2: Placements of targets in the study with degrees and centimeters from the middle line.**

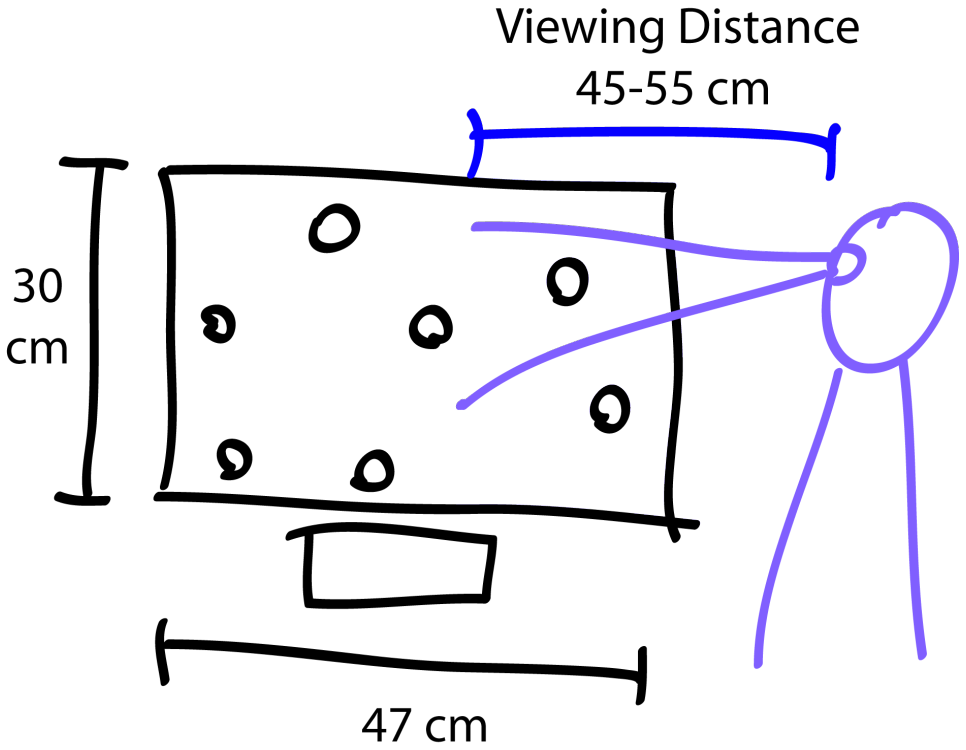

**Figure 3.3: The setup used for the experiment**

### 3.1.3 Results

We found no significant difference in acquisition rate between the cues for both patients and students. However, for the two neglect patients we saw strong indications of *color* being worse than the remaining cues. The grand mean of acquisition rate for *color* was 90.4% while the neglect patients had a mean acquisition rate of 66.7%. *Expansion* had a grand mean of 92.1% with 80% for neglect patients. *Luminance* had a grand mean of 94.8% with 84.2% for neglect patients. Finally, *peripheral motion* had a grand mean of 94.8% with 85.8% for neglect patients. There was no significant difference in acquisition rate between students (99.7%) and patients (89%).

Performing ANOVA on the individual reaction times revealed that there was no significant interaction between target distance from the centerline and visual cueing method. See Figure 3.4 for the reaction times for the individual participants. For patients, cueing type was found to be significant (F(3,92), $p \ll 0.001$). The Tukey's HSD test revealed that *expansion* had a significantly worse reaction time, 1.56 seconds on average, compared to the grand mean of 1.24 seconds for the other cues (23.9 % faster). For the students, a Friedman test revealed a significant difference between cue types ($X^2(3) = 14.25$, $p < 0.01$). The post hoc comparison showed the same pattern, with *expansion* as the significantly slower cue, with an average reaction time of 0.71 seconds compared to the grand mean of 0.61 for the other cues (15.2 % faster).

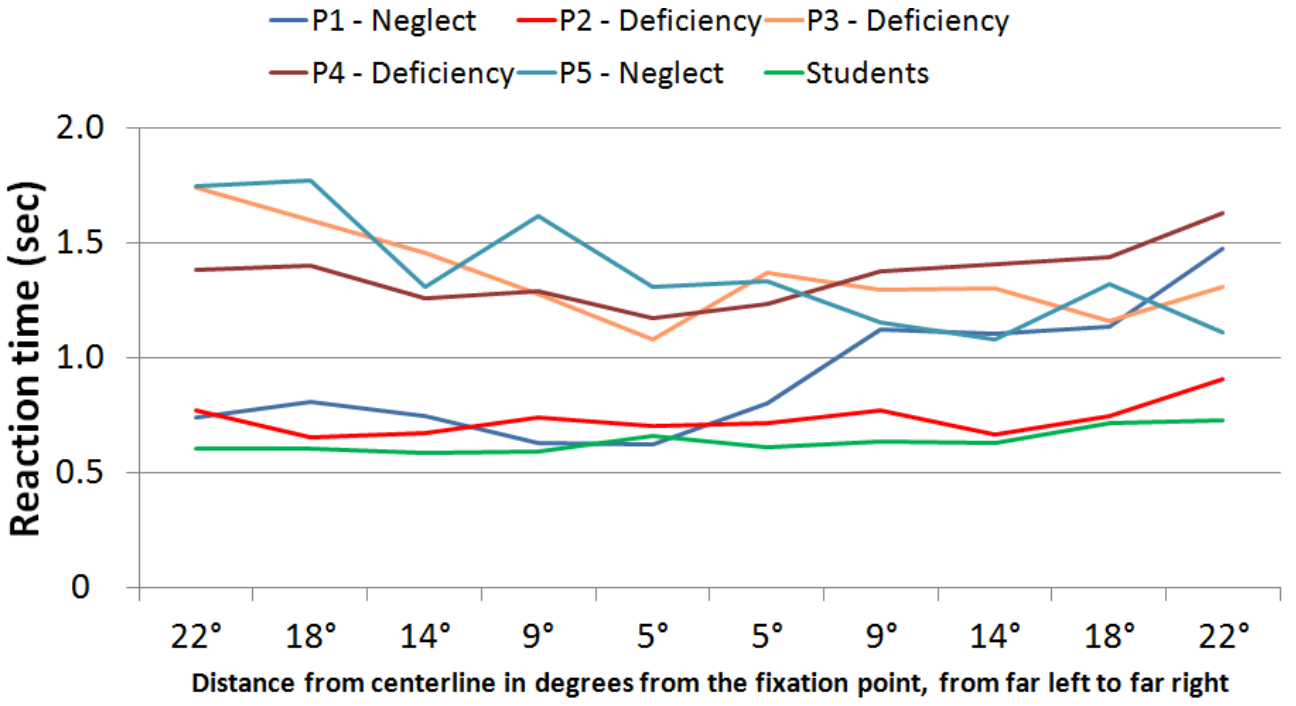

**Figure 3.4: Reaction time performance of the individual participants for each distance from the middle line. Distances correspond to those shown on Figure 3.**

Overall, three of the cues were close in reaction time, with *expansion* having roughly 0.3 seconds longer reaction time on average. However, the reaction time indicates that as the distance from the center increases so does the reaction time. Comparing the students to the patients, an independent t-test revealed that the student group ($M$ = 1.32, $SD$ = 0.81) was significantly faster than the patient group ($M$ = 0.64, $SD$ = 0.24) (t(5.1) = 3.57, $p < 0.05$). From the eye tracking data, we saw that participants followed the instructions and fixated on the center cube over 80 % of the time ($M$ = 81.3 %, $SD$ = 13.6).

### 3.1.4 Partial Conclusion

Our results show that *expansion* is significantly worse in terms of reaction time for all participant groups. The other cues are not significant and are within 7 ms of each other. This corresponds to our expectations, as all of the cues are strong visual cues [16]. Reaction time tended to increase linearly with target distance from the centerline; however, there was no significant effect of it. Similarly, we found that age increased the overall reaction time; this corresponds to findings by Kirchner et al. [8]. For the young neglect patient, we saw that the reaction time increased from that of the similar aged, to levels similar to the older participants. The results also indicate that *peripheral motion* had a positive influence on reaction time, compared to a simple color change. The preference of the patients was overall spread between *luminance* and *peripheral motion*. In regard to target acquisition, we found that the *expansion* and *color* had the worst performance. This is especially the case for people with visual neglect.

In an interesting event, the younger neglect patient preferred *color* as a visual cue. However, this cue caused the patient to miss the most targets. The patient thought he had acquired all targets.

## 3.2 Second Study

For the second study, we examined the impact of endogenous cues on trail making tasks. We tested three cues based on arrows (denoted *breadcrumbs*, *magnetic field*, and *compass*), two on lines (denoted *static line* and *dynamic line*) and two wedges (denoted *static wedge* and *dynamic wedge*) based on the wedge cue by Gustafson et al. [6]. Figure 3.5 illustrates the seven designs. *Breadcrumbs* and *magnetic field* became progressively easier, as previous targets become arrows. This makes previous targets easily distinguishable in the shape feature channel.

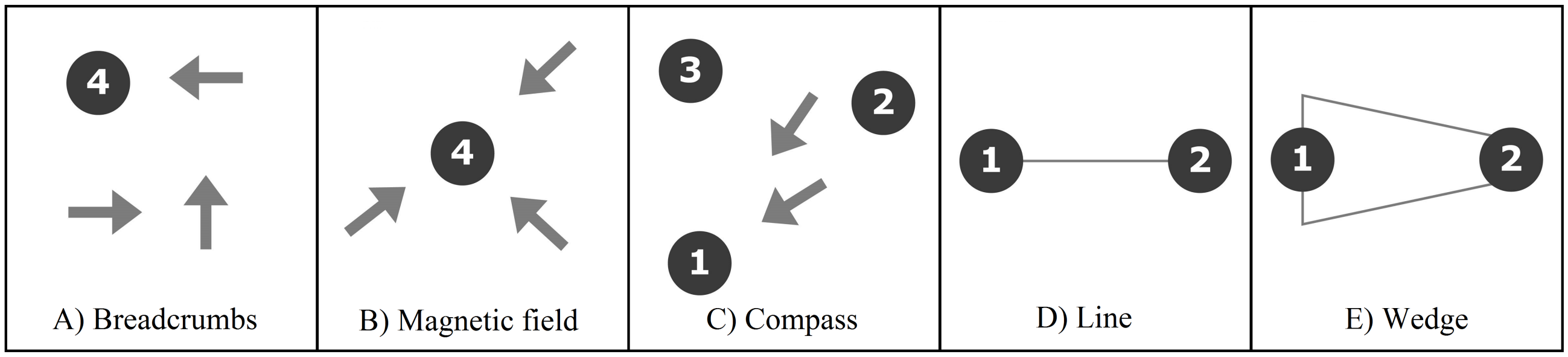

**Figure 3.5: The visual cues used during the second study. (A) *Breadcrumbs*, when selected targets transition into arrows that point towards the next target. (B) *Magnetic field*, same principle as (A) but all arrows point towards the same target. (C) *Compass*, two arrows located above and below the center vertical line and points toward the next target. (D) Line, the line either starts at the previous target (static) or follows the gaze (dynamic). (E) Wedge, the base of the wedge either starts at the previous target (static) or follows the gaze (dynamic).**

The second study only had patients with an attention deficit, but a visual neglect patient from the third study agreed to participate in the second study. As a result, the data from this patient is separate from the others.

### *3.2.1 Design and Procedure*

The study used a within subject design, with the exogenous cues as the main independent variable. The dependent variables were target acquisition time, completion time of a task and acquisition rate.

The study consisted of eight trail making tasks, seven with different endogenous cues and one without cues (denoted *no cue*). Before the experiment, the participant went through a training session. The training session consisted of simpler trail making tasks in contrast to the more complex versions in the experiment. The training session continued until participants felt confident with the tasks and all cues. The trail making tasks involved selecting numeral targets in an increasing order, as seen in Figure 3.6. The selection of targets happened with mouse input by clicking on a target. On a correct selection, the target would change color to green until a new selection occurred. An incorrect selection was marked with red; this would disappear when the participant selected the correct target. In order to complete a task, participants needed to select targets one through nine. The entire experiment took approximately 15 minutes.

### *3.2.2 Apparatus and Materials*

We conducted the experiment on a 22" screen with a resolution of 1600 x 1050 and used an ambidextrous mouse set to 1000 DPI with mouse acceleration disabled. We utilized a Tobii X120 for tracking the scan paths and a Tobii Eye-X to control the dynamic cues. Participants had a viewing distance of 45-55 cm to the screen.

### *3.2.3 Participants*

Five voluntary participants from a neurorehabilitation center, in the age range of 24 to 71 ($M = 50.20$, $SD = 18.30$) along with three students from the local university campus, in the age range of 24 to 27 ($M = 25.33$, $SD = 1.53$). All participants were right-handed and had experience with a mouse as an input device.

### *3.2.4 Expectations*

The expectation is that the wedge and line cues perform better, as they directly create a link to the targets, rather than just providing a direction. We also expect that the performance of *breadcrumbs* decreases as the trial progresses, due to the number of arrows pointing in different directions.

### *3.2.5 Results*

The average completion times for all participants reveals that there is a significant difference between cue types ($X^2(7) = 14.29$, $p < 0.05$). The Friedman post hoc test reveals that *no cue* is significantly slower than all cueing techniques, except for *compass*, which in turn is only significantly slower than the *static line*. The Friedman test on the acquisition times for each cue shows a similar result. However, here *magnetic field* and *static line* share the position of being significantly faster than *no cue* and *compass*. We checked the acquisition time for each cue for correlation with the distances between the targets. Here, we found that only two methods had significant correlation at *p*

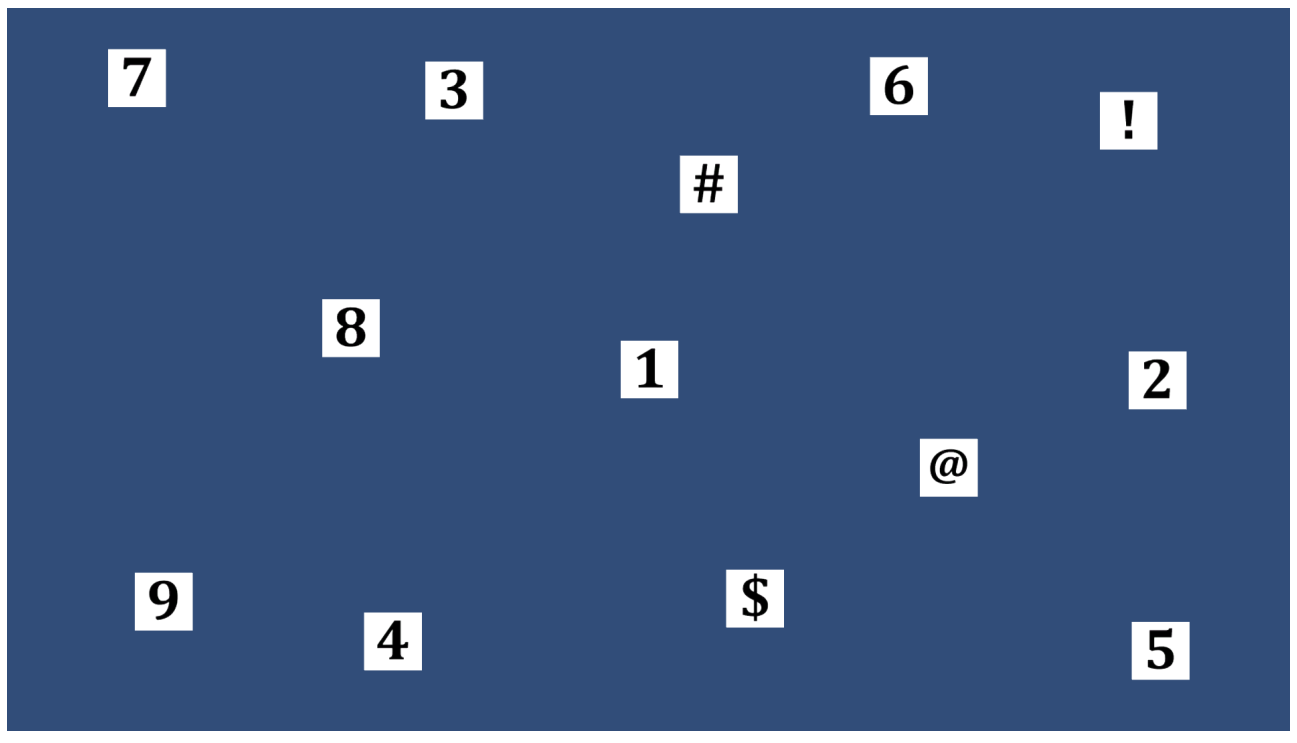

**Figure 3.6: A trail making task with nine targets (numerals) and four distracters (symbols). The first target always appeared at the center.**

< 0.05, namely the *static line* and the *breadcrumbs*. The participants made no errors.

In the first study, we saw a difference in the performance of patients and students. Therefore, we examined if this was the case again. Figure 3.7 shows the completion times normalized to match acquisition times for each cue. Here, we see that the completion times for students are overall lower. The student's results do not deviate as much between methods, except for *no cue*, which has an overall higher completion time. An independent t-test revealed that students ($M$ = 9.9, $SD$ = 2.1) completed the tasks significantly faster than attention deficit patients ($M$ = 16.3, $SD$ = 4.2) ($t(5.85) = 4.53$, $p < 0.01$). A potentially influencing factor in acquisition times is the presence of distracting targets in path from the current position to the next target. An dependent t-test on the acquisition times revealed that the presence of distracting targets ($M$ = 1.86, $SD$ = 0.98) significantly increased acquisition times compared to no distracters ($M$ = 0.93, $SD$ = 0.79) ($t(7) = -3.39$, $p < 0.05$). Figure 3.8 illustrates which visual cues are the most affected by the presence of distracting targets. Here, we see that there is little for most visual methods, except for *compass* and the *static wedge*.

We found no effect of the order in which each set of cues appeared. We also found no significant correlation between the performance of students and patients for each visual cue. Significant effects of target order were found for some of the cues, however post hoc tests revealed only few individual differences, and there were no patterns indicating acquisition time decreasing as users progressed through the test.

### 3.2.6 Eye Tracking

We analyzed the eye tracking by examining the scan paths for each participant using Tobii Studio. To find patterns, we superimposed the scan paths of each cue on each other.

The scan paths revealed an overarching pattern for the three arrow cues (*breadcrumbs*, *magnetic field*, and *compass*);

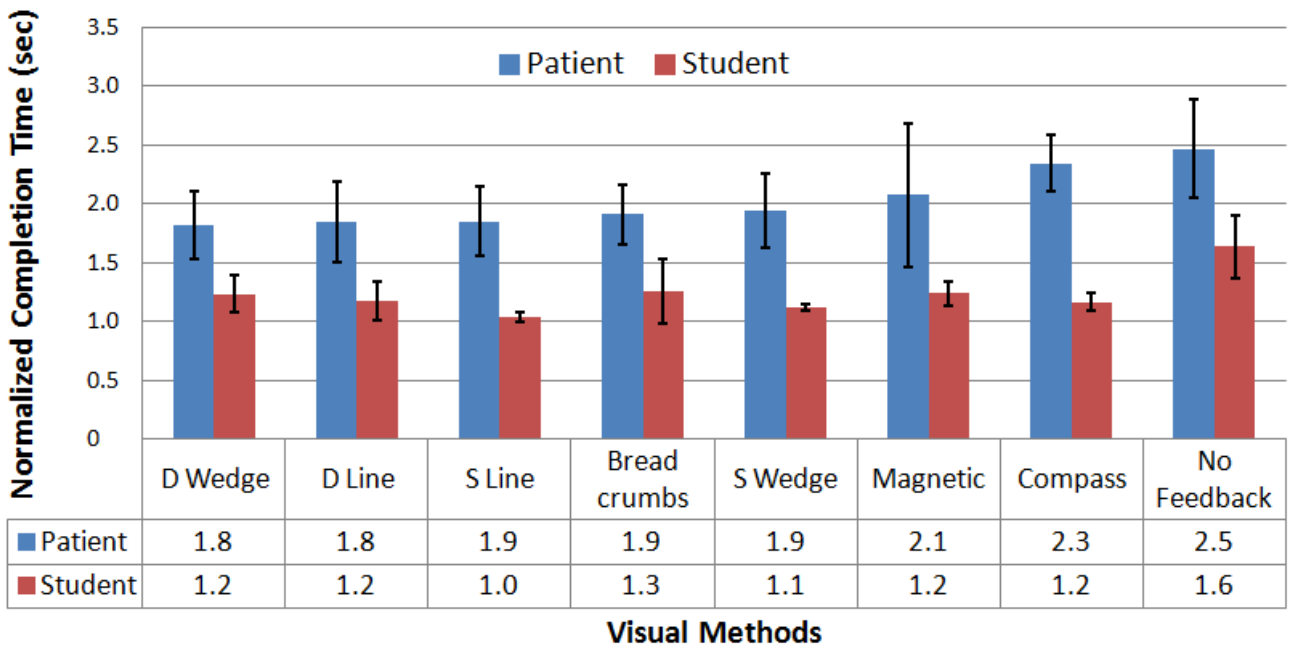

| | D Wedge | D Line | S Line | Bread crumbs | S Wedge | Magnetic | Compass | No Feedback |
|---|---|---|---|---|---|---|---|---|
| Patient | 1.8 | 1.8 | 1.9 | 1.9 | 1.9 | 2.1 | 2.3 | 2.5 |
| Student | 1.2 | 1.2 | 1.0 | 1.3 | 1.1 | 1.2 | 1.2 | 1.6 |

**Figure 3.7: Average normalized completion time for all visual methods illustrated for the patient group and the student group. The patient group has higher variance in their performance for almost each case compared to the student group.**

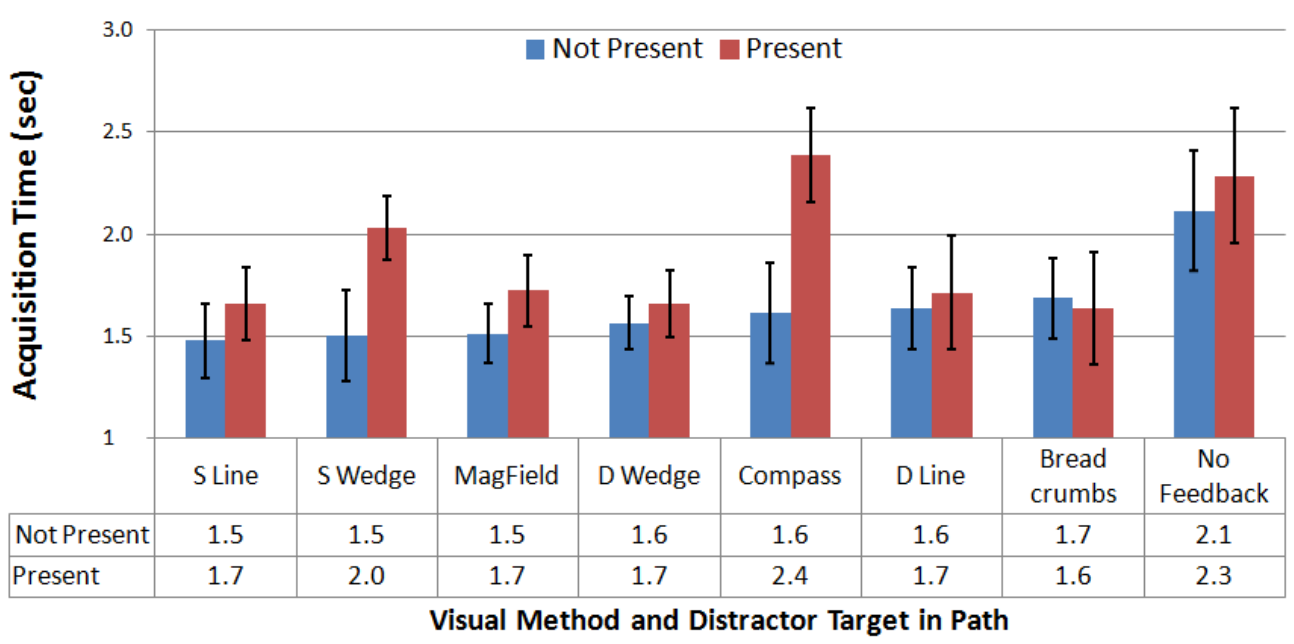

| | S Line | S Wedge | MagField | D Wedge | Compass | D Line | Bread crumbs | No Feedback |
|---|---|---|---|---|---|---|---|---|
| Not Present | 1.5 | 1.5 | 1.5 | 1.6 | 1.6 | 1.6 | 1.7 | 2.1 |
| Present | 1.7 | 2.0 | 1.7 | 1.7 | 2.4 | 1.7 | 1.6 | 2.3 |

**Figure 3.8: Average acquisition time for all visual methods illustrated for the cases where a distracter or other number was present in the path to the next target.**

participants would saccade towards targets in the general direction of the pointing arrows. The scan paths from these techniques form a cone shape, with the tip starting at the arrow and the base expanding towards the direction of the arrow. For the *compass*, the cone behavior was mirrored in the sense that base would start at the arrows with the tip expanding towards the target. See Figure 3.9 for a visualization of these cones. The number of saccades needed to find the target with these techniques seems to increase linearly with the number of targets encapsulated by the cones. In a sense, these techniques only showed direction and still required a visual search to acquire the target.

The scan paths for the *static wedge* and *static line* also shared a pattern; participants would follow the line until reaching the target. For the wedge, they would follow either the top or lower wedge line. The number of saccades needed to find the target for these techniques, increased linearly with the number of targets touching or within around one cm to the line(s).

### 3.2.7 User Preference

For user preference, we ranked the participants' opinions on a five-point scale. Patients preferred the *static line*, while students preferred the *magnetic field*. See Figure 3.10 for the user ratings.

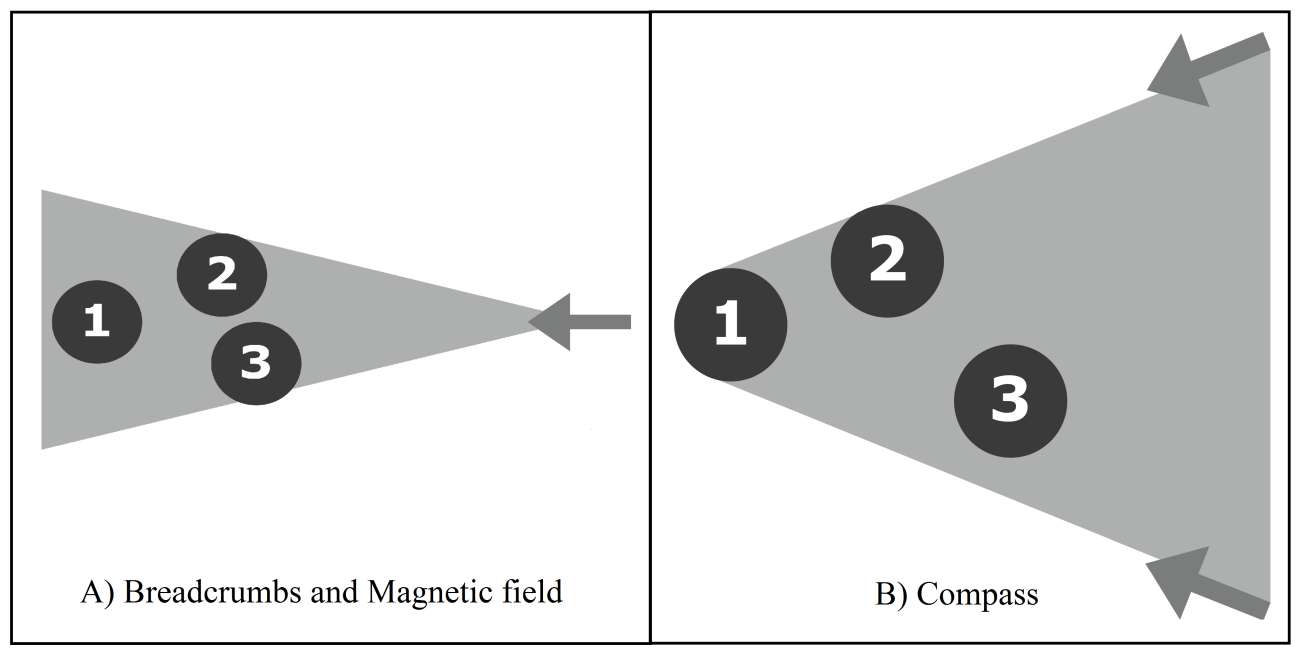

**Figure 3.9: The gray area indicates the cone, in which participants looked to find the target**

### 3.2.8 Neglect Patient

We found no significant differences between the methods; however, we saw indications of the *static wedge* and *breadcrumbs* being better. For *breadcrumbs*, acquisitions became faster with progression, while the *static wedge* was more consistent. These findings are in line with our expectations, as *breadcrumbs* decreases in difficulty while *static line* has a linear difficulty.

Regarding the eye tracking, the visual neglect patient showed an interesting behavior for both the line and wedge technique, as the patient would follow the line into the neglected area but would slow down upon reaching the area and proceed with small saccades along the line.

### 3.2.9 Partial Conclusion

In terms of average acquisition time and completion time, the line cues have an edge over the other methods. The *dynamic line* had the lowest average completion time for patients, while *static line* had the lowest average completion times for students. *No cue* performed worst, with *compass* having a slightly poorer performance than the other cues. Furthermore, the test showed that the students had lower completion times, compared to the attention deficit patients. The variance of acquisition times was greater for patients, compared to the students that had greater similarity in acquisition times.

The *breadcrumbs*, *compass*, and *magnetic field* cues, all points towards targets, however, we observed that it was hard for participants to find the exact location of the target based on the direction of the arrows alone. Participants deviated towards targets in the general direction of the arrow; in a sense, these cues only showed direction and not location. In particular, the *compass* cue suffered because of this, as the nature of the *breadcrumbs* and *magnetic field* reduces the number of targets gradually as participants select them. The wedge cues were also susceptible to these problems, as there was a larger area in which other targets could appear within compared to the line cues. These findings are in line with our expectations, as the line and wedge cues create a link between targets compared to the others that only show a direction.

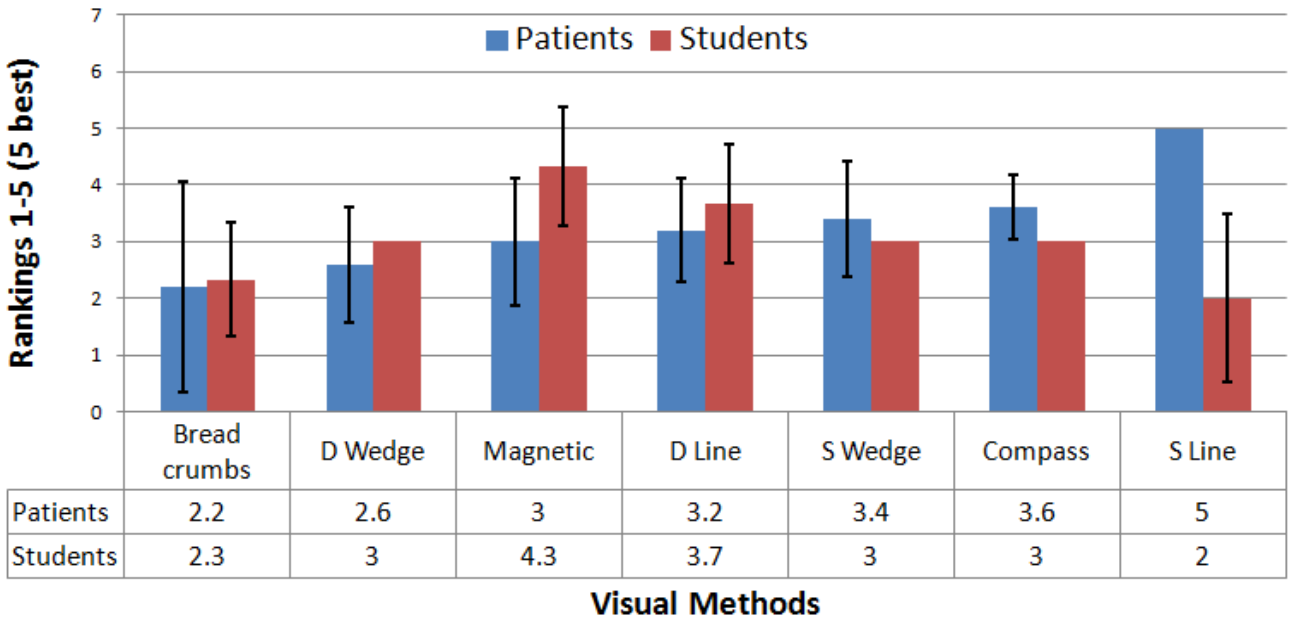

**Figure 3.10: Average methods rankings. Preference rankings are based on the participants stated opinion about each method. The best and worst method was ranked five and one respectively; methods in-between was rated based on positivity or negativity of statements.**

Patients tended to favor the *static line*, while performing better with the *dynamic line*, while students tended to dislike the *static line*, but still overall perform better with it. Several patients commented on the dynamic methods as being non-calm or messy, however the additional movement may have helped them draw their attention to the right target. Further, the students pointed out that the *static line* implicated two targets, while the dynamic only implicated one.

Based on the results, we have chosen to use *static line* and *static wedge* for the third study. The dynamic versions are interesting, however, with the current setup, the eye tracker used for controlling the *dynamic line* and *dynamic wedge* interferes with the eye tracking system used to track the scan paths. We will decrease the width of the wedge from eight to four degrees, to decrease the span of the wedge. This will furthermore help decrease confusion, as there is less space for potential distracters. As an additional benefit, this will help the wedge take on a shape closer to that of a directional arrow.

## 3.3 Third Study

The goal of this study was to test the impact of endogenous cues paired with exogenous cues on the ability to find targets. Based on the previous studies, we have chosen eight visual cue designs. From the first study, we have the *luminance*, where a target will pulsate between being in a lit and unlit state, and *peripheral motion*, where a bar will move vertically in the outer edge of the screen depending on the direction of the next target. From the second study, we use the *static line* and *static wedge*. The visual cues new to this study are the combinations of these exogenous and endogenous cues (denoted *line with luminance*, *line with peripheral motion*, *wedge with luminance* and *wedge with peripheral motion*).

### 3.3.1 Procedure

The study consisted of nine tasks, one for each type of visual cue and one without visual cues. A task consisted of a modified trail-making task with 20 numbered targets and six symbol targets as additional distracters. The users had to select the targets with a mouse pointer and not connect the targets. This enabled the difficulty to remain the same, as only the previous target had a highlight and not all the previous targets. Selecting the correct target turned the target green, while an incorrect selection would cause the selected target to turn red. The cues for the next target would appear once the user selected a correct target, along with the previous cue disappearing. The first target always appeared at the center of the screen, so users had a point to return to for the beginning of each task.

We introduced each visual cue type through a training session that contained a smaller version of the experiment with only five targets. The training session continued until the participant felt confident with the different cues and tasks. The duration of the experiment was approximately 20 minutes.

### 3.3.2 Design

The study used a within subject design, with visual cues as the main independent variable. In addition, we examined the distance between targets, number of distracters in the path to the current target, target position, movement direction, target order, and the order in which the cues appeared. The dependent variables were the acquisition time from target to target, and the task completion time. We used eye tracking as an additional diagnostic tool to investigate the scan paths of the participants. Based on the second study, we wanted to examine the effect of multiple distracters between the starting point and target. Therefore, the sequence of targets was positioned such that each cue type had instances of zero to six distracters present. We ensured that the participant would always have to go across both the horizontal and vertical middle line at least 10 times to cover more of the visual field. Similarly, the travel distances approximated one of four travel distances, the distances being 16, 22, 28, and 34 cm. For analysis, we divided the target position into four overall categories, inner and outer left and right side. See Figure 3.11.

### 3.3.3 Apparatus and Materials

The study had a similar setup and apparatus to that of the second study.

### 3.3.4 Participants

The study had two visual neglect patients aged 58 and 72; the older patient had severe neglect. Five students in the age range 23 to 28 ($M$ = 25, $SD$ = 2.8) and five middle aged citizens in the age range 58 to 70 ($M$ = 63.6, $SD$ = 4.6). All participants voluntarily participated in the study. All participants were right handed and had experience with the mouse as an input device.

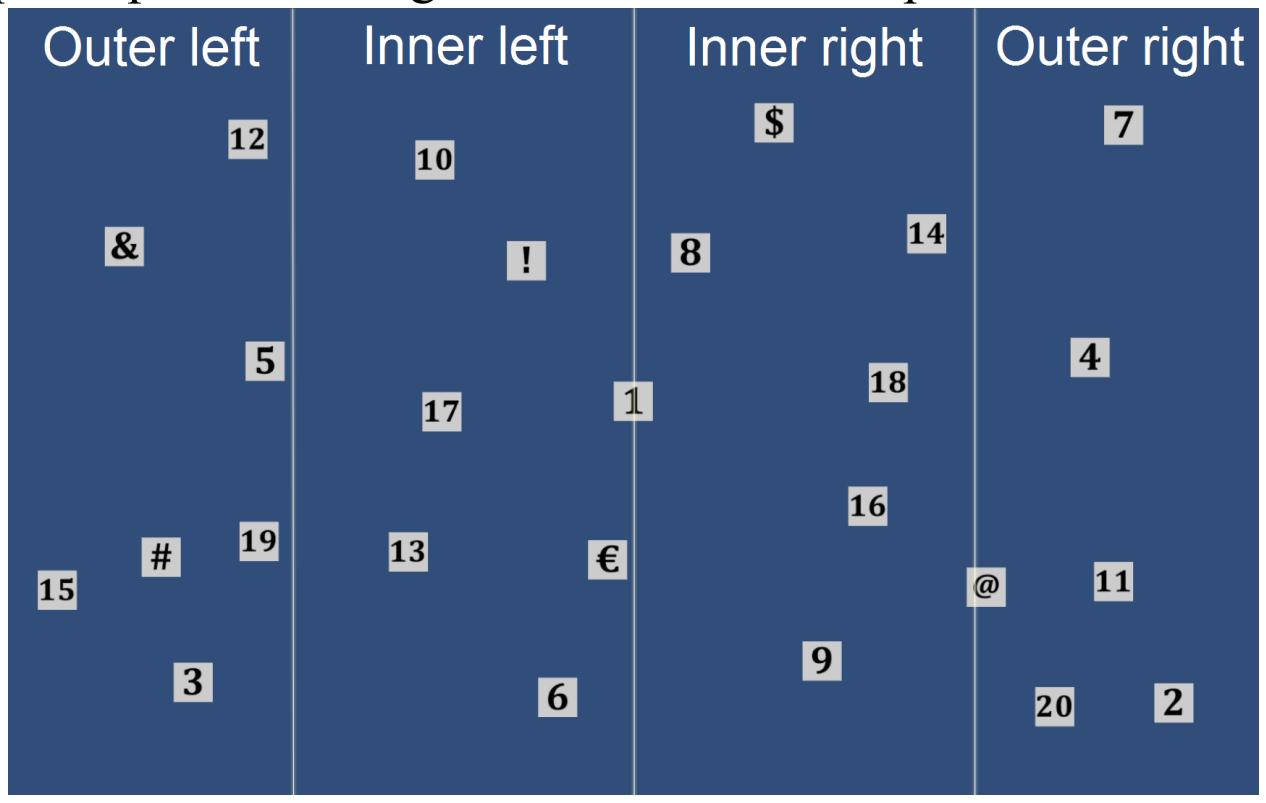

**Figure 3.11: Target positions was split into four columns based on their horizontal position, outer left, inner left, inner right, and outer right.**

### 3.3.5 Expectations

Based on our previous studies, we expect that the endogenous cues (*line* and *wedge*) will perform better as they provide a bridge into the neglected area. We further expect that the cue combinations help visual neglect patients pay attention to the neglected area and thereby provide a performance improvement. For the standalone *peripheral motion*, we expect that it will be better as an addition rather than by itself as the cue only shows direction. For participants with normal sight, we do not expect there to be much difference between the cues, except for *peripheral motion*. We expect that *peripheral motion* requires a longer visual search, as the cue only shows direction of the target based on the previous target. Lastly, we expect all cues to be better than *no cue*.

### 3.3.6 Results

We examined each of the participant groups to see which methods worked best for each group. We examined the neglect patient's data individually, as one had severe neglect that caused the patient to interact differently with the cues (denoted neglect patient and severe neglect patient). We also terminated the study prematurely for the severe neglect patient, due to the patient having severe difficulties completing the tasks. This means that there is no data for *line with peripheral motion* and *wedge with peripheral motion*.

We used ANOVA along with the Tukey's HSD post hoc for all comparisons. We examined the additional independent variables of target order, method presentation order, travel distance, distracting targets, movement direction, and target position, where not mentioned, no significant effect was found. Comparisons use acquisition time as dependent variable, unless otherwise stated.

For the patient with neglect, the results showed a large effect of cue type ($F(8,162) = 6.8$, $p << 0.01$, $r = 0.51$). All line and wedge methods were significantly faster than the remaining conditions. The average acquisition times show that the *wedge* and *wedge with luminance* had the best performance, but not significantly, see Table 3.1.

Movement direction showed to be highly significant ($F(2,168) = 34.2$, $p <<0.01$, $r = 0.51$), with leftwards taking 3.6 seconds longer than rightwards movements (128 % slower). Similarly target position, had a large effect ($F(3,167) = 19.6$, $p << 0.01$, $r = 0.51$), with the outer left targets taking 2.2 seconds longer than the inner left targets, which in turn takes 2.1 seconds longer than targets in the right side. Finally, there was a small effect of distance between targets ($F(3,167) = 3.6$, $p < 0.05$, $r = 0.24$). The average time used on target distances of 28 cm (5.44 seconds, 78 % slower) and 34 cm (5.14 seconds, 67 %) are significantly higher than the target distance of 16 cm (3.05 seconds). We found no effect of the order of cues or the order of targets.

**Table 3.1: Average acquisition times for each visual cue and participant group. Wedge is shortened to W, line to L, Peripheral Motion to PM, and Luminance to Lum. The severe neglect patient is missing data entries for two cues.**

| Cue | Neglect Left - Right | | Severe Neglect Left - Right | | Middle aged | Students |
|---|---|---|---|---|---|---|
| W Lum | 2.8 | 1.7 | 14.1 | 5.2 | 1.7 | 1 |
| Wedge | 3.5 | 1. 9 | 10.7 | 6.7 | 1.7 | 1 |
| L Lum | 3.7 | 2.1 | 11.8 | 5.2 | 1.7 | 1.1 |
| L PM | 4 | 1.9 | - | - | 1.7 | 1 |
| W PM | 5.7 | 2.2 | - | - | 1.7 | 1.1 |
| Line | 6.6 | 2.2 | 21.9 | 5.1 | 1.6 | 1 |
| PM | 8.2 | 4.8 | 17.7 | 9 | 3.4 | 2.3 |
| Lum | 6.5 | 4.6 | 15.8 | 7.9 | 2.4 | 1.5 |
| No cue | 15.2 | 3.8 | 18.2 | 7.7 | 3.7 | 2.6 |

The patient with severe neglect often required additional help to find targets within the neglected area. We found no significant effect of the method used. However, there is trend of the line and wedge cues being faster. We saw a large effect of interaction direction ($F(2,130) = 23.21$, $p << 0.01$, $r = 0.51$), with leftward movements taking 8.8 seconds longer on average (126.4 % slower). Matching this, we also saw an effect of target position ($F(3,129) = 21.4$, $p << 0.01$, $r = 0.58$), with acquisition time in the inner left area being 5.7 seconds higher (83.8 % slower) than acquisition time in the right side, and additional 5.8 seconds higher (46.4 % slower) in the outer left area.

The middle-aged participants had a large effect of cue type ($F(8,846) = 112$, $p << 0.01$, $r = 0.59$). All wedge and line cues had a significantly faster acquisition time (1.7 seconds) compared to *luminance* (2.4 seconds, 34.1 % slower), *peripheral motion* (3.4 seconds, 66.7 % slower) and *no cue* (3.7 seconds, 74.1 % slower). See Table 1 for the individual acquisition times. We observed a small effect of target position ($F(3,851) = 6.1$, $p << 0.01$, $r = 0.15$). Inner targets were 2.5-3 seconds faster to acquire than the outer targets in each side. Additionally, we found a small effect of distracters in the path ($F(6,848) = 4.2$, $p << 0.01$, $r = 0.17$), a time increase of 0.6-1.1 seconds was found between the presences of zero to one and four to six distracters.

The student group also had a large effect of cue type ($F(8,846) = 112$, $p << 0.01$, $r = 0.71$). All feedback types using line and wedge were quicker (1.1 seconds) than *luminance* (1.5 seconds, 36.4 % increase) and *peripheral motion* (2.3 seconds, 109.1 % increase), which in turn were better than *no cue* (2.6 seconds, 136.4 % increase over line and wedge), see Table 1.

Examining overall completion time for participant groups, an effect of group was present ($F(2,2026)$, $p << 0.01$, $r = 0.66$). The student group overall completed the tasks significantly faster (26.6 seconds on average) than the middle-aged group (42.2 seconds on average, 58.7 % slower than the student group), who in turn were significantly faster than the patients (144 seconds on average, 241.2 % slower than the middle-aged group), see Figure 3.12.

### *3.3.7 Eye Tracking*

The analysis followed the same procedure outlined in the second study.

The scan path for the wedge and line cues had the same pattern as in the second study, even with the increased number of targets. The scan paths also show that the addition of the exogenous cues (*luminance* and *peripheral motion*) had no influence on the scan paths. However, the neglected patients deviate from this pattern. The patient with visual neglect followed a pattern similar to those of students and middle aged but slowed down when the wedge and line would extend into the neglected area. The eye motions were reduced to small saccades along the line or wedge. If a distracter would appear on the line within the neglected area, the patient would stop, look back at the previous target then continue along the line past the distracter. This pattern would repeat until the patient found the target. However, for the combined *line with luminance* cue, the patient would continue along the line without looking back at the previous target. The severe neglect patient had trouble following all the line and wedge cues into the neglected area. As a result, the tasks within the neglected area required prolong visual search. However, the patient showed no distinct patterns during visual search.

For *luminance*, students found it without visually searching for it, whereas the middle-aged group had to briefly visually search for the cue. The further the targets were in the periphery, the longer they had to search for it. The patients with visual neglect had difficulties finding the cue when in the neglected area and required a prolonged visual search to find it but would otherwise behave as the middle-aged group in the non-neglected area.

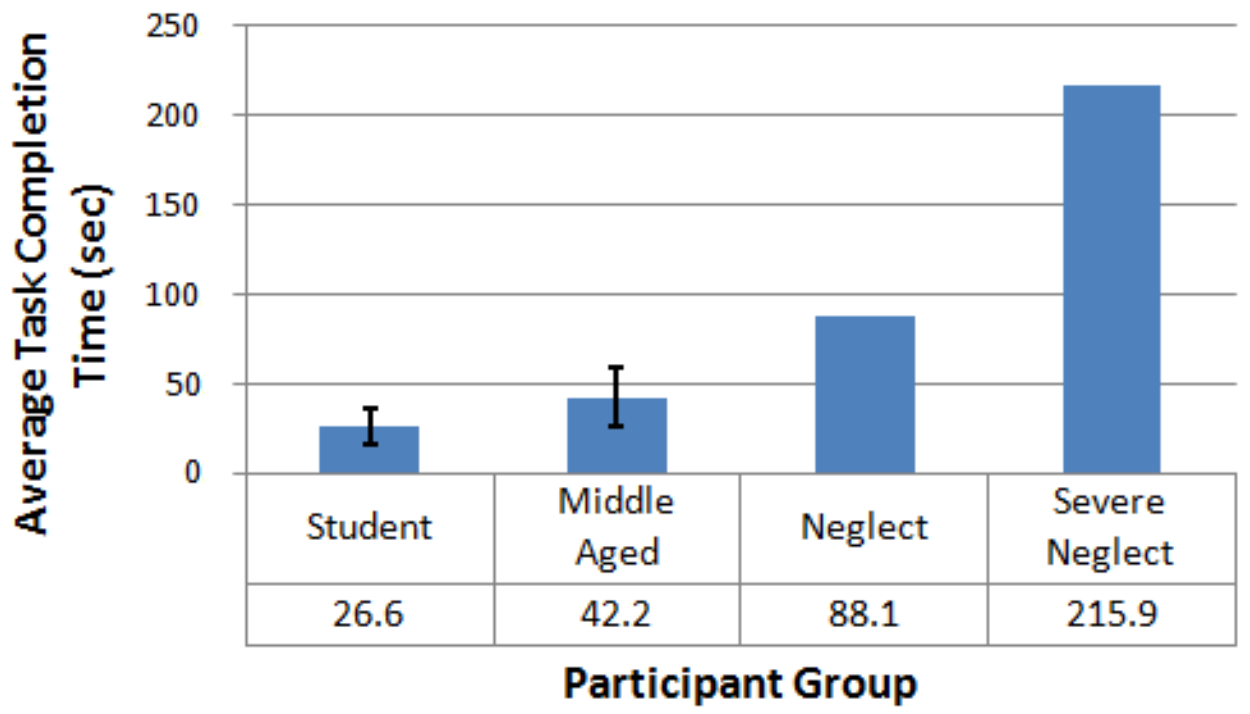

**Figure 3.12: Average task completion time for each group.**

Scan paths for *peripheral motion* showed that the student group and middle-aged group still had to visually search for the target, however they only searched in the direction of the bar. This means that the cue reduced the search area. For the patient with neglect, the bar attracts attention and drag the gaze towards the bar instead of searching for the target. This cue had no influence for the patient with severe neglect.

The scan paths of the patient with neglect also indicate that for the exogenous cues and *no cue*, the patient would overcompensate for the neglect by focusing the visual search in that area. In this sense, the patient neglected the non-neglected side. In comparison, the patient with severe neglect only began searching within the neglected area after having searched the non-neglected area.

#### *3.3.8 Preference*

The prevalent opinion among the middle-aged participant group was a preference for the *line* and *wedge* specifically, while not caring about the exogenous additions to these methods. Exogenous were worse than the line and wedge methods according to the middle-aged participants. The students mostly preferred *wedge*, while stating that cues were almost equally good. However, *wedge* added the benefit of removing all doubt about which end the correct one was. According to the students, the light pulsations in *luminance* were a bit too slow to be of real use, while *peripheral motion* alone did not accomplish much. The patient with visual neglect stated a preference for the line methods, and *luminance*. The patient described that the addition of exogenous cues improved upon *line*. The wedge cues, while doing the job, caused him to expect there to be something more to the task, compared to the line cues. The severe neglect preferred the line methods and mentioned that both the wedge and luminance methods were good, but not as good. The severe neglect patient also stated that luminance was a good addition to *line* and *wedge*.

## 4. DISCUSSION

In the first experiment, we saw that *expansion* had the worst performance, with a similar performance for the other cues. The experiment also suggests that the further targets are in the periphery the harder targets are for patients with visual neglect to see. The younger neglect patient had a similar response time to students for the non-neglected area, while the response time approached that of those with an attention disorder as targets extended into the periphery in the neglected area. For the older neglect patient, the response time matched that of the attention disorder patients regardless of position. However, for both neglect patients, the target acquisition rate decreased as the targets extended further into the neglected periphery. The younger neglect patients were in the same age range as the students, while the older patients were in the age range of those with an attention disorder. This suggests a relation between acquisition time and age, instead of time and disorder.

From the second experiment, we saw that the line and wedge cues performed better or similar to the other cues. The *breadcrumbs* and *magnetic field* cues performed on par with the line and wedge cues; however, these cues also transformed all previous targets whereas the other designs only showed the last target. This potentially favored these designs over the others, but it also shows the strength of the line and wedge designs despite having a disadvantage they still performed better on average.

The third experiment confirmed the strength of the line and wedge designs as they outperformed the exogenous cues for all participant groups. The results also showed that combining the exogenous cues with the line and wedge had no influence on acquisition and completion time along with scan paths for both the students and middle-aged group. These findings are in contrast to Ware [15], who reported that the more distinctive features an object has, the easier it is to find. However, it could also suggest that the cue needed to be even more distinct. With *luminance*, this can be an issue as the distinctiveness of the cue depends on the amount of natural light. To make it more distinctive would require dimming the surroundings, which is not always viable. However, the additions to the wedge and line cues did not hinter the cues suggesting that combining them does not negatively influence acquisition time. This could also be the result of a ceiling effect, meaning that cues cannot improve beyond human reaction time and motor control.

Participants from the student group also expressed that the pulsation speed and total luminance level was too low. Adjusting these values could improve *luminance*, especially for subjects with reduced peripheral vision.

## 5. CONCLUSION AND FUTURE WORK

We evaluated exogenous and endogenous cues for visual neglect patients, based on acquisition time and acquisition rate. Furthermore, we compared the results to students and middle-aged participants. In the first study, we evaluated exogenous cues and found that a luminance pulsation and peripheral bar cue performed better than a color change and expansion cue. As expected from previous research the cues performances were not significant [16]. The second study evaluated endogenous cues and found that linking targets together with a either a wedge or a line improves acquisition time, compared to cues that only show direction. The findings from the second study are based on attention deficit patients rather than visual neglect patients. In the third study, we combined the two best exogenous cues for visual neglect patients with the two best endogenous cues for attention deficit patients. The results show that the endogenous cues help participants find the targets more quickly, thereby decreasing acquisition time. Based on the results from the healthy patients there did not seem to be a performance benefit from combining the endogenous and exogenous cues. However, one participant with neglect performed better with the combination of cues compared to endogenous cues alone. The patient with severe neglect did not seem to benefit as much from the cues, indicating that cues might not work identically for each neglect patient.

The next iteration of this study will be to expand upon the visual cues and further iterate upon the existing cues. A potential iteration would be to see how visual neglect patients behave if the line curved around distracters. Furthermore, previous research [4, 13] have shown that combining visual and auditory cues improve performance for visual neglect patients. Therefore, future iterations should investigate how combining these cues influence the search patterns of visual neglect patients.

## 6. ACKNOWLEDGEMENTS

We would like to express our sincere gratitude to the neurorehabilitation centers in Skive, Hammel and Brønderslev. We would like to thank all volunteers who participated in the experiments.